\documentclass[sigconf]{acmart}

\usepackage{amsmath}
\usepackage{balance}
\usepackage{booktabs} 
\usepackage{graphicx}
\usepackage{multirow}
\usepackage{rotating}
\usepackage{soul}
\usepackage{url}
\usepackage[utf8]{inputenc}
\usepackage{color}
\usepackage{colortbl}
\usepackage{subcaption}
\usepackage[export]{adjustbox}
\usepackage{stackengine}
\usepackage{amsfonts}
\usepackage{amssymb}
\usepackage{amsthm}
\usepackage[normalem]{ulem}
\usepackage{array}

\newtheorem{property}{Property}

\newif\ifdraft
\drafttrue

\newif\ifspaceneeded
\spaceneededtrue

\newcommand{\removeifspaceneeded}[1]{\ifspaceneeded{}\else{#1}\fi}

\copyrightyear{2019} 
\acmYear{2019} 
\setcopyright{acmlicensed}
\acmConference[SIGIR '19]{Proceedings of the 42nd International ACM SIGIR Conference on Research and Development in Information Retrieval}{July 21--25, 2019}{Paris, France}
\acmBooktitle{Proceedings of the 42nd International ACM SIGIR Conference on Research and Development in Information Retrieval (SIGIR '19), July 21--25, 2019, Paris, France}
\acmPrice{15.00}
\acmDOI{10.1145/3331184.3331308}
\acmISBN{978-1-4503-6172-9/19/07}

\begin{document}

\title[Evaluating Multiple-Option Lists in Chatbots and Mobile
Search]{Evaluating Variable-Length Multiple-Option Lists \\ in
Chatbots and Mobile Search}

\author{Pepa Atanasova}
\affiliation{University of Copenhagen, Copenhagen, Denmark}
\author{Georgi Karadzhov, Yasen Kiprov}
\affiliation{SiteGround Hosting EOOD, Sofia, Bulgaria}
\author{Preslav Nakov}
\affiliation{Qatar Computing Research Institute, HBKU, Doha, Qatar}
\author{Fabrizio Sebastiani}
\affiliation{Istituto di Scienza e Tecnologie dell'Informazione, Consiglio Nazionale delle Ricerche, 56124 Pisa, Italy}


%

\renewcommand{\shortauthors}{Atanasova et. al.}

\begin{abstract}

  In recent years, the proliferation of smart mobile devices has lead
  to the gradual integration of search functionality within mobile
  platforms. This has created an incentive to move away from the ``ten
  blue links'' metaphor, as mobile users are less likely to click on
  them, expecting to get the answer directly from the snippets. In
  turn, this has revived the interest in Question Answering.  Then,
  along came chatbots, conversational systems, and messaging
  platforms, where the user needs could be better served with the system
  asking follow-up questions in order to better understand the user's
  intent.  While typically a user would expect a single response at
  any utterance, a system could also return multiple options for the
  user to select from, based on different system understandings of the
  user's intent. However, this possibility should not be overused, as this practice
  could confuse and/or annoy the user.  How to produce good
  variable-length lists, given the conflicting objectives of staying
  short while maximizing the likelihood of having a correct answer
  included in the list, is an underexplored problem. It is also
  unclear how to evaluate a system that tries to do that. Here we aim
  to bridge this gap. In particular, we define some necessary and some
  optional properties that an evaluation measure fit for this purpose
  should have. We further show that existing evaluation measures from
  the IR tradition are not entirely suitable for this setup, and we
  propose novel evaluation measures that address it satisfactorily.
\end{abstract}


\maketitle


\section{Introduction}
\label{sec:intro}

\noindent Mobile devices have emerged as an essential
and integral part of our lives. 
Yet, the limited size of mobile screens, and the consequently
reduced amount of displayed information, have brought about
new challenges in terms of user experience. The first step towards
addressing them is to depart
from the ``ten blue links'' metaphor and to actually understand the
user's information needs~\cite{Baeza-Yates:2011qd,Russell-Rose:2012yg}.
Moreover, chatbots have been introduced as a way to include a
follow-up interaction with the user \cite{folstad2017chatbots}.

Most chatbots in actual
use~\cite{Coucke:2018mq,Bocklisch:2017jh,Burtsev:2018hw,Williams:2015ht}
share a common characteristic: they are capable of handling different
types of user needs, usually called \emph{intents}.  
As the chatbot might engage in an elaborate 
series of actions and utterances
to fulfill the user's 
intent, a high-accuracy intent detection module becomes a
crucial component of such systems.  Yet, as the number of intents
being handled grows, the accuracy of the intent classifier tends to
decrease, with reported $F_{1}$ values down to 0.73 and even to
0.52~\cite{Braun:2017am,Coucke:2018mq}, depending on the domain and
the number of intents considered.

In order to mitigate intent misclassification, personal assistants can use 
strategies such as continue with the most likely intent, ask for confirmation,
return a list of possible intents, or repeat the question. Previous research has found that users prefer a list of the most likely intents, but also noted that this ``complicates with clutter; unnatural; more reading'' \cite{ashktorab2019resilient}, i.e.,~the list should be concise.

To this end, toolkits such as the IBM Watson
Assistant\footnote{\url{http://console.bluemix.net/docs/services/assistant/dialog-runtime.html}}
and Oracle's Digital Assistant\footnote{\url{https://bit.ly/2XbW1Do}}
provide functionality for defining confidence thresholds, which allow
more candidate intents to be displayed when the model is not
confident.  This means less typing by the user and faster
narrowing down the user's request \cite{ashktorab2019resilient}.

The examples we present below show that going in the wrong direction
might have very negative consequences.  Instead, the system could
present a list of highly likely options and then leave it to the user
to select the correct one, e.g.,

\vspace{1ex}
\noindent
\begin{small}
  \begin{tabular}{ll|lp{32mm}}
    \bf User: & \emph{My credit card is toast.} & \bf User: & \emph{I  want to cancel it today}\\
    Bot: & What do you want to do? & Bot: & What do you want to cancel?\\ 
    Bot: & $\blacktriangleright$Replace a broken card. & Bot: &  $\blacktriangleright$ account\\
    Bot: &  $\blacktriangleright$Report a stolen card. & Bot: &  $\blacktriangleright$ card\\
              & & Bot: &  $\blacktriangleright$ last transaction \\
  \end{tabular}
\end{small}

\noindent The system should be careful, though, not to suggest too many options,
as this could confuse and/or annoy the user
\cite{chaves2019chatbot}. Moreover, it should try to make sure the
list contains a good suggestion. Depending on the presentation mode,
the order in which the options are presented might or might not
matter.

\noindent Current research and development for chatbots moves in the direction
of open-domain task-oriented systems, with an ever-increasing number
of intents, which makes variable-length lists much more
common. However, the evaluation of such systems remains an
underexplored problem, and existing measures from the IR tradition do
not fit this setup well enough. Therefore, we define a set of
properties that an evaluation measure should satisfy in order to
optimize two conflicting objectives simultaneously,
i.e.,~(\emph{i})~reduce the size of the list, and (\emph{ii})~maximize
the likelihood that a good option is indeed in the list.  We further
propose evaluation measures that satisfy all these desiderata.  While
here we focus on chatbots, most of the arguments we present apply to
mobile search, too.


\section{Assumptions and Desiderata}
\label{sec:propertiesfortopn}

\noindent Our goal is to evaluate systems that, given a user question,
try to understand the underlying intent and to answer with a suitable response. 
We assume that the question expresses a single user intent and therefore the system can return a single correct response. \footnote{We leave the case of multiple possible correct answers for future work.}
We further assume that the system always returns a
non-empty list of responses,\footnote{We assume a special
\emph{default} intent with a default answer to cover the case when the
system cannot understand the intent.} and that different responses
correspond to different intents. We represent response lists as
sequences of symbols from $\{c,w\}$, where $c$ stands for a correct
response and $w$ stands for a wrong one.
Finally, we assume that the position of the correct 
answer 
(if returned) in the list may or may not matter, depending on the context. 
In other words, we cater for
the fact that in some applications the results should be considered a
plain unordered \textit{set}, while in some others they should form a
\textit{ranked list}.

Next, we define a set of properties that an evaluation measure
$M$ for variable-length lists should satisfy. Given two response
lists $\mathbf{r}_{1}$ and $\mathbf{r}_{2}$ 
for the same user question, a property specifies which
one $M$ should give a higher score. Note that we take $M$ to be a measure of
accuracy, and not of error, i.e., higher values of $M$ are better. We use $r_{ij}\in\{c,w\}$ to refer to the response item at the
$j$-th position of $\mathbf{r}_{i}$. 
We further define a function
$\#_{s}(\mathbf{r_{i}})\equiv|\{r_{ij}\in \mathbf{r}_{i}|r_{ij}=s\}|$
that, given a response list $\mathbf{r}_{i}$, returns the number of
responses of type $s$ (where $s\in\{c,w\}$) that $\mathbf{r}_{i}$
contains. Moreover, we define a function
$p(\mathbf{r_{i}})\equiv\sum_{r_{ij}=c}\frac{1}{rank(r_{ij})}$ that,
given a response list $\mathbf{r_{i}}$, 
returns the reciprocal rank of the correct response, or 0 if no correct response was returned. 
Finally, we define a re-scaling function
$s(x, newMAX)\equiv x*newMAX$, which re-scales $x$ from the
range [0, 1] to the range [0, newMAX].  We use $|\mathbf{r}_{i}|$
for the \emph{length} of $\mathbf{r}_{i}$, and the symbol $>$ to
express preference between two lists.

\begin{property}\label{ax:Correctness}
  (\textbf{Correctness}) If
  $\#_{c}(\mathbf{r_{1}})>\#_{c}(\mathbf{r_{2}})$,\\ then
  $M(\mathbf{r}_{1}) > M(\mathbf{r}_{2})$.  \hfill \ \qed
\end{property}

\noindent This property states that a response list that contains
a correct response should be preferred to one that does not.

\begin{property}\label{ax:Confidence} 
  (\textbf{Confidence}) If
  $\#_{c}(\mathbf{r_{1}})=\#_{c}(\mathbf{r_{2}})$\\ and
  $\#_{w}(\mathbf{r_{1}})<\#_{w}(\mathbf{r_{2}})$, then
  $M(\mathbf{r}_{1}) > M(\mathbf{r}_{2})$.  \hfill \ \qed
 
\end{property}

\noindent This property states that, if two lists contain the same number of correct responses 
(can be 0 or 1),
the list with fewer wrong responses is preferable. 
The aim is to limit the length of the response list.

\begin{property}\label{ax:Priority} 
  (\textbf{Priority}) If
  $\#_{c}(\mathbf{r_{1}})=\#_{c}(\mathbf{r_{2}})$ and
  $\#_{w}(\mathbf{r_{1}})=\#_{w}(\mathbf{r_{2}})$ and
  $p(\mathbf{r_{1}}) \geq p(\mathbf{r_{2}})$, then
  $M(\mathbf{r}_{1}) \geq M(\mathbf{r}_{2})$.  \hfill \ \qed
\end{property}

\noindent This property states that if two lists
both contain a correct response, then the list where the correct response is ranked higher should be preferred.

We view \textbf{\textit{Correctness}} and \textbf{\textit{Confidence}}
as mandatory properties for all our measures, and
\textbf{\textit{Priority}} as an optional one, depending on
whether the results from the chatbot application are presented as an
unordered set or as a ranked list.

\removeifspaceneeded{Note that the introduced properties follow a certain priority
order when used to identify a preference between result lists. At
first, we use the \textbf{\textit{Correctness}} property and only if
it fails to identify a preference, we will apply the
\textbf{\textit{Confidence}} property. Then, if
\textbf{\textit{Confidence}} also fails, we will use the
\textbf{\textit{Priority}} property. Finally, if the
\textbf{\textit{Priority}} property also fails, we will consider the
evaluation scores of the lists to be equal.}


\section{Evaluation Measures}
\label{sec:topnevaluationmet}

\begin{table*}[h!]
  \centering \small
  \begin{tabular}{|l||c|r|r|r||r|r|r|r|r|r|r|r|r|r|}
    \hline
    &  \multicolumn{4}{c||}{\bf Unranked Retrieval} & \multicolumn{10}{c|}{\bf Ranked Retrieval}\\
    \midrule
    \multicolumn{1}{|c||}{Result list} & \multicolumn{1}{c|}{Gold} & \multicolumn{1}{c}{$F_{1}$} & \multicolumn{1}{|c}{$F_{1}^s$} & \multicolumn{1}{|c||}{\textit{LAR}} & \multicolumn{1}{c|}{Gold} & \multicolumn{1}{c}{\textit{AP}} & \multicolumn{1}{|c}{\textit{AP}$^L$} & \multicolumn{1}{|c}{\textit{AP}$^s$} & \multicolumn{1}{|c}{\textit{RR}} & \multicolumn{1}{|c}{\textit{nDCG}} & \multicolumn{1}{|c}{\textit{nDCG$^L$}} &
                                                                                                                                                                                                                                                                                                                                                                                                                                       \multicolumn{1}{|c}{\textit{RBP}} &
                                                                                                                                                                                                                                                                                                                                                                                                                                                                           \multicolumn{1}{|c}{\textit{RBP$^L$}} &
                                                                                                                                                                                                                                                                                                                                                                                                                                                                                                                   \multicolumn{1}{|c|}{\textit{OLAR}} \\ 
    \midrule
    c & 1 & 1.00 & 1.00 & 1.00 & 1 & 1.00 & 1.00 & 1.00 & 1.00 & 1.00 & 1.00 & 0.50 & 1.00 & 1.000 \\ 
    cw & 2 & 0.67 & 0.80 & 0.75 & 2 & (*) 1.00 & 0.83 & 0.83 & (*) 1.00 & (*) 1.00 & 0.92 & (*) 0.50 & 0.75 & 0.756 \\ 
    wc & 2 & 0.67 & 0.80 & 0.75 & 3 & 0.50 & 0.58 & 0.58 & 0.50 & 0.63 & 0.69 & 0.25 & 0.50 & 0.744 \\ 
    cww & 4 & 0.50 & 0.67 & 0.67 & 4 & (*) 1.00 & (*) 0.75 & (*) 0.75 & (*) 1.00 & (*) 1.00 & (*) 0.88 & (*) 0.50 & (*) 0.63 & 0.675 \\ 
    wcw & 4 & 0.50 & 0.67 & 0.67 & 5 & (*) 0.50 & 0.50 & 0.50 & (*) 0.50 & (*) 0.63 & 0.65 & (*) 0.25 & 0.38 & 0.663 \\ 
    wwc & 4 & 0.50 & 0.67 & 0.67 & 6 & 0.33 & 0.42 & 0.42 & 0.33 & 0.50 & 0.57 & 0.13 & 0.25 & 0.659 \\ 
    cwww & 7 & 0.40 & 0.57 & 0.63 & 7 & (*) 1.00 & (*) 0.70 & (*) 0.70 & (*) 1.00 & (*) 1.00 & (*) 0.85 & (*) 0.50 & (*) 0.56 & 0.634 \\ 
    wcww & 7 & 0.40 & 0.57 & 0.63 & 8 & (*) 0.50 & (*) 0.45 & (*) 0.45 & (*) 0.50 & (*) 0.63 & (*) 0.62 & (*) 0.25 & (*) 0.31 & 0.622 \\ 
    wwcw & 7 & 0.40 & 0.57 & 0.63 & 9 & (*) 0.33 & 0.37 & 0.37 & (*) 0.33 & (*) 0.50 & 0.54 & (*) 0.13 & 0.19 & 0.618 \\ 
    wwwc & 7 & 0.40 & 0.57 & 0.63 & 10 & 0.25 & 0.33 & 0.33 & 0.25 & 0.43 & 0.50 & 0.06 & 0.13 & 0.616 \\ 
    cwwww & 11 & 0.33 & 0.50 & 0.60 & 11 & (*) 1.00 & (*) 0.67 & (*) 0.67 & (*) 1.00 & (*) 1.00 & (*) 0.83 & (*) 0.50 & (*) 0.53 & 0.610 \\ 
    wcwww & 11 & 0.33 & 0.50 & 0.60 & 12 & (*) 0.50 & (*) 0.42 & (*) 0.42 & (*) 0.50 & (*) 0.63 & (*) 0.61 & (*) 0.25 & (*) 0.28 & 0.598 \\ 
    wwcww & 11 & 0.33 & 0.50 & 0.60 & 13 & (*) 0.33 & (*) 0.33 & (*) 0.33 & (*) 0.33 & (*) 0.50 & (*) 0.52 & (*) 0.13 & (*) 0.16 & 0.594 \\ 
    wwwcw & 11 & 0.33 & 0.50 & 0.60 & 14 & (*) 0.25 & 0.29 & 0.29 & (*) 0.25 & (*) 0.43 & 0.48 & (*) 0.06 & 0.09 & 0.591 \\ 
    wwwwc & 11 & 0.33 & 0.50 & 0.60 & 15 & 0.20 & 0.27 & 0.27 & 0.20 & 0.39 & 0.46 & 0.03 & 0.06 & 0.590 \\ 
    w & 16 & 0.00 & ($\bigtriangleup$) 0.50 & 0.50 & 16 & 0.00 & 0.00 & 0.25 & 0.00 & 0.00 & 0.00 & 0.00 & 0.00 & 0.488 \\ 
    ww & 17 & (*) 0.00 & 0.40 & 0.25 & 17 & (*) 0.00 & (*) 0.00 & 0.17 & (*) 0.00 & (*) 0.00 & (*) 0.00 & (*) 0.00 & (*) 0.00 & 0.244 \\ 
    www & 18 & (*) 0.00 & 0.33 & 0.17 & 18 & (*) 0.00 & (*) 0.00 & 0.13 & (*) 0.00 & (*) 0.00 & (*) 0.00 & (*) 0.00 & (*) 0.00 & 0.163 \\ 
    wwww & 19 & (*) 0.00 & 0.29 & 0.13 & 19 & (*) 0.00 & (*) 0.00 & 0.10 & (*) 0.00 & (*) 0.00 & (*) 0.00 & (*) 0.00 & (*) 0.00 & 0.122 \\ 
    wwwww & 20 & (*) 0.00 & 0.25 & 0.10 & 20 & (*) 0.00 & (*) 0.00 & 0.08 & (*) 0.00 & (*) 0.00 & (*) 0.00 & (*) 0.00 & (*) 0.00 & 0.098 \\    \hline
    \multicolumn{2}{|l|}{\bf \textit{Correctness}}  & \multicolumn{1}{c|}{Yes}  & \multicolumn{1}{c|}{No} & \multicolumn{1}{c||}{Yes}  & & \multicolumn{1}{c|}{Yes}  & \multicolumn{1}{c|}{Yes} & \multicolumn{1}{c|}{Yes}  & \multicolumn{1}{c|}{Yes}  & \multicolumn{1}{c|}{Yes} & \multicolumn{1}{c|}{Yes} & \multicolumn{1}{c|}{Yes} & \multicolumn{1}{c|}{Yes} & \multicolumn{1}{c|}{Yes} \\ 
    \multicolumn{2}{|l|}{\bf \textit{Confidence}}& \multicolumn{1}{c|}{No} & \multicolumn{1}{c|}{Yes}  & \multicolumn{1}{c||}{Yes}  & & \multicolumn{1}{c|}{No} & \multicolumn{1}{c|}{No} & \multicolumn{1}{c|}{No} & \multicolumn{1}{c|}{No} & \multicolumn{1}{c|}{No} & \multicolumn{1}{c|}{No} & \multicolumn{1}{c|}{No} & \multicolumn{1}{c|}{No} & \multicolumn{1}{c|}{Yes} \\ 
    \multicolumn{2}{|l|}{\bf \textit{Priority}}& \multicolumn{1}{c|}{No} & \multicolumn{1}{c|}{No} & \multicolumn{1}{c||}{No} & & \multicolumn{1}{c|}{Yes}  & \multicolumn{1}{c|}{Yes}  & \multicolumn{1}{c|}{Yes}  & \multicolumn{1}{c|}{Yes}  & \multicolumn{1}{c|}{Yes} & \multicolumn{1}{c|}{Yes}  & \multicolumn{1}{c|}{Yes} & \multicolumn{1}{c|}{Yes} & \multicolumn{1}{c|}{Yes} \\
    \hline
    \multicolumn{2}{|l|}{\textit{\textbf{Kendall's Tau}}} & 0.970 & 0.985 & 1 & & 0.746 & 0.827 & 0.857 & 0.746 & 0.746 & 0.811 & 0.746 & 0.811 & 1 \\
    \multicolumn{2}{|l|}{\textit{\textbf{Spearman correlation}}} & 0.992 & 0.994 & 1 &  & 0.855 & 0.926 & 0.934 & 0.855 & 0.855 & 0.918 & 0.855 & 0.918 & 1 \\
    \hline
  \end{tabular}
  \caption{Comparison of evaluation measures.  The 1st column indicates
  all response lists with up to 5 responses.  ``Gold'' columns contain
  the ideal ranking of these lists according to our properties, while
  the other columns contain the evaluation scores from the
  corresponding measures.  We designate inconsistencies in the
  rank order w.r.t.\ the gold order with an asterix (*) when due to
  violation of \textbf{\textit{Confidence}}, and with a triangle
  ($\bigtriangleup$) when due to violation of \textbf{\textit{Correctness}}.
  The compliance of the measures with the properties is
  indicated at the bottom of the table, where the correlation of the
  ranking with the gold ranking is also shown.}\label{tab:results}
\vspace{-7.5mm}
\end{table*}
\noindent In Table~\ref{tab:results}, we present an evaluation of 
existing information retrieval measures w.r.t.\ the introduced
properties.
The
response lists are ranked according to the properties at hand, and
together with the priority between the properties, we obtain a unique
``gold ranking.''

Next, we compare the evaluation scores and the resulting rank
order of the various measures with respect to the ``gold ranking.'' To this end,
we estimate Kendall's Tau and Spearman's rank correlation between the
two and we indicate the errors in the rank positions. We also provide
information about the properties that each of the measures satisfies or violates.

\textbf{Existing Measures for Unranked Retrieval.}  Considering
evaluation of unordered sets, one obvious candidate is $F_{1}$.  We
can see in Table~\ref{tab:results} that $F_{1}$ is successful at
rewarding the presence of the correct answer (due to the recall
component) and, usually, at minimizing the length of the response list
(due to the precision component). Unfortunately, its value is always 0
when there 
is no correct response,
Thus, it fails to satisfy \textbf{\textit{Confidence}}.

We try to solve the problem by smoothing $F_{1}$ (denoted as $F_{1}^{s}$), which we obtain by appending an
extra correct response at the end of each list.  The
resulting measure does not suffer from the above problems of $F_{1}$,
but fails to distinguish 
between a list consisting of a single wrong
response and lists with one correct and four wrong responses, thus
failing to satisfy \textbf{\textit{Correctness}}.

\textbf{Existing Measures for Ranked Retrieval.} A natural candidate
measure for ranked retrieval is \emph{Average Precision}
(\textit{AP}). It computes precision after each relevant response,
thus satisfying both \textbf{\textit{Correctness}} and
\textbf{\textit{Priority}}. However, it disregards the number of the
returned irrelevant responses and even stops computing at the last
relevant response, ignoring all the subsequent irrelevant responses,
and thus it fails to satisfy \textbf{\textit{Confidence}}.

In order to allow the length of the response list to influence the evaluation score, it has been proposed \cite{Liu:2016qo} to append a terminal
response \textbf{t} at the end of each response list, which is called the $\mathit{AP}^{L}$ measure.  $\mathit{AP}^{L}$
manages to alleviate some of the ranking problems observed
in \textit{AP}, but still fails to satisfy  
\textbf{\textit{Confidence}} in some cases.

Another way to make \textit{AP} penalize wrong responses at the
end of the list is to use smoothing (denoted
as $\mathit{AP}^{s}$). Now, the more wrong responses we return
at the end of the list, the lower the precision at the last recall
level will get. As a result, $\mathit{AP}^{s}$ manages to reduce further the number
of errors in the ranking with respect to the gold order. Nevertheless, $\mathit{AP}^{s}$ still
does not satisfy \textbf{\textit{Confidence}}  in
cases when the result lists have different numbers of wrong results and
different positions of the correct responses, as in \textit{wcw} and
\textit{cwww}. This is due to $\mathit{AP}^{s}$ failing to apply the
properties in the correct order, i.e., applying \textbf{\textit{Priority}}
before \textbf{\textit{Confidence}}.

\emph{Reciprocal Rank} (\textit{RR}) is a popular measure for ranked
retrieval, which accounts for the position of the 1st correct
response, disregarding any following irrelevant responses. 
\textit{AP}
is equivalent to \textit{RR} in the case of a single correct response
(and so is \textit{DCG}).

Furthermore, although \emph{normalized Discounted Cumulative
Gain} (\textit{nDCG}) is designed specifically for the evaluation of
different relevance scores, we study its performance on our ``gold
ranking.'' However, it does not penalize wrong responses, violating
\textbf{\textit{Confidence}}. Using the technique proposed in
\cite{Liu:2016qo}, we end up with $\mathit{nDCG}^{L}$, which manages
to reduce the number of ranking errors, but still violates 
\textbf{\textit{Confidence}}  in some cases.

\cite{moffat2008rank} proposed  \textit{Rank-Biased Precision}
\textit{(RBP)}, which models a user that decides to continue reading the next item in the response list with probability $p$. As in \cite{Liu:2016qo}, we set $p$=0.5. \textit{RBP}
struggles with the same problems as \textit{RR} and violates 
\textbf{\textit{Confidence}}. Even if we apply the technique
from \cite{Liu:2016qo}, the ranking produced by
$\mathit{RBP}^{L}$ contains a lot of errors compared to the ``gold
ranking.''

\textbf{New Measures.}  The existing measures we have discussed are
suitable for optimizing the number of correct
responses and their positions. Most of them fulfill
\textbf{\textit{Correctness}} and \textbf{\textit{Priority}}, but
struggle with \textbf{\textit{Confidence}}. This is not surprising as the IR tradition (except for recent work like
\cite{Albahem:2018yq,Liu:2016qo}) is concerned with the rank positions
of the relevant documents, not with the length of
the response list, which is conceptually infinite. Before
\cite{Albahem:2018yq,Liu:2016qo}, this length had never been a
parameter in any of the proposed measures. 

\noindent Smoothing lists by
appending an additional correct response is beneficial but not
enough to achieve perfect correlation with the ``gold ranking,'' as the
Kendall's Tau and the Spearman rank correlation scores indicate.
In order to bridge this gap, we introduce a new measure, \emph{Length-Aware Recall} (\textit{LAR}), which operates on a list of responses
$\mathbf{r}_{n}$ and gives preference to lists with fewer
negatives:
\begin{equation}
  \mathit{LAR} = \frac{R(\mathbf{r}_{n}) +  \frac{1}{|\mathbf{r}_{n}|}}{2}
\end{equation}
The new measure first computes the recall $R(\mathbf{r}_{n})$ of the returned list. Then, it includes the confidence of the system about the correct result
expressed by the reciprocal value of the list's length, i.e.,~the
confidence decreases when the number of returned responses
increases. Taking the mean of the two scores, we create an intuitive
score of both recall and confidence. The possibility of having zero
values for recall makes arithmetic mean preferable to harmonic mean,
also used to combine evaluation criteria.

As \textit{LAR} satisfies both \textbf{\textit{Correctness}} and
\textbf{\textit{Confidence}}, it can be used to evaluate variable-length lists by modeling the true positive rate and the optimal response length jointly. Moreover, it is
perfectly correlated with the ``gold ranking'' in the unordered
scenario.

However, \textit{LAR} does not satisfy
the \textbf{\textit{Priority}} property, which makes it unfit for scenarios
where order does matter. In order to fix this issue, we propose an extension of  \textit{LAR} that includes an additional third term for the rank of the correct response. 

\noindent This fix gives rise to
\textit{Ordered Length-Aware Recall} (OLAR):
\begin{equation}
  \mathit{OLAR} = \frac{R(\mathbf{r}_{n}) +  \frac{1}{|\mathbf{r}_{n}|} + s(p(\mathbf{r}_{n}),\mu)}{2+\mu}
\end{equation}
The third term in the above equation accounts for \textbf{\textit{Priority}}, and it is larger when the rank of the correct item moves
lower in the list. We rescale it because of the
priority order that we have defined for the properties -- it should not contribute to the
score more than the \textbf{\textit{Confidence}} term. Given that $(\frac{1}{max(|r_{i}|)-1} - \frac{1}{max(|r_{i}|)}) = 0.05 $ 
is the smallest difference between two
\textbf{\textit{Priority}} scores of lists with length up to 5, we re-scale it in [0, $\mu$], where $\mu= 0.05-\lambda$ and $\lambda=0.001$ is an insignificantly small
number.

To sum up, we introduced two measures, which are in a perfect
correlation with the two ``gold rankings.'' The measures consist of
separate terms, accounting for different properties, which makes them easily interpretable and
extensible for specific needs.


\section{Related Work}
\label{sec:relatedwork}

\textbf{Related properties of evaluation measures.}  As we define a
set of properties that need to be satisfied by an evaluation measure
for variable-length output, our work is closely related to the
properties of truncated evaluation measures introduced in
\cite{Albahem:2018yq}. Their \textit{Relevance Monotonicity} property
is similar to our \textit{Correctness}, except that
\textit{Correctness} imposes strict monotonicity. The
\textit{Irrelevance Monotonicity} property is similar to our \textit{Confidence}, but it discounts for irrelevant documents at the end of the list only.

\cite{moffat2013seven} defined seven properties for effectiveness
measures. The relevant properties, which our new evaluation measures
also satisfy are \textit{Completeness}, \textit{Realisability},
\textit{Localisation} and \textit{Boundedness}. However, their
\textit{Monotonicity} property contradicts our \textit{Confidence}
property because it states that adding documents, which are not
relevant, to the end of the list increases the score.

\cite{Amigo:2018bc,Sebastiani:2015zl} also conducted an
``axiomatic'' analysis of IR-related evaluation
measures. Our properties \textit{Confidence} and \textit{Priority} are
akin to the ones discussed in~\cite{Amigo:2013ai}, but the latter are
used in a different setup, i.e., for document retrieval, clustering, and filtering.

\textbf{Related evaluation measures.}  In Section~\ref{sec:topnevaluationmet}, we discussed and evaluated the most
relevant evaluation measures for both unranked (precision, recall, $F_{1}$, and
smoothed $F_{1}$) and ranked retrieval
(\textit{nDCG}~\cite{Jarvelin:2000vh}, \textit{RR}, \textit{MAP}, smoothed
\textit{MAP}, \textit{RBP} \cite{Liu:2016qo}).  We found that they
were unable to penalize the wrong responses according to the ``gold
order,'' thus violating  \textit{Confidence}.

Apart from these measures, \cite{penas2011simple} 
introduced the \textit{c@1} measure, a modification of accuracy,
suited for systems that may not return responses. However, their
approach still does not penalize the number of returned wrong
responses at the end of the list. Furthermore, \cite{penas2011simple,
voorhees2001overview, sakai2004new} assumed that a system can return
an empty result list, which tackles the problem when the request does
not have a correct response. However, we assume
that the system should return at least one result, even if it is a
default fallback intent.

Another relevant field of research analyzes the likelihood that the
user will continue exploring the response list based on different
signals - time-biased gain \cite{smucker2012modeling}, length of the
snippet \cite{sakai2013summaries}, and information foraging
\cite{azzopardi2018measuring}.  However, in mobile search and chatbot
platforms, the presented information is already minimized, and thus we aim
to reduce the length of the returned results instead.


\section{Conclusion and Future Work}
\label{sec:conclusion}

\noindent We have studied the problem of evaluating variable-length
response lists for chatbots, given the conflicting objectives of
keeping these lists short while maximizing the likelihood of having a
correct response in the list. In particular, we argued for three
properties
that such an evaluation measure should have. We further
showed that existing evaluation measures from the IR tradition do not
satisfy all of them. Then, we proposed novel measures that are
especially tailored for the described scenarios, and satisfy all
properties.

We plan to extend this work to the context in which more than one
correct answer might exist, since a long and complex input question
may contain multiple intents~\cite{Xu:2013pt}. This would also be of
interest for mobile retrieval in general, where results may need to be
truncated due to limited screen space.

%

\bibliographystyle{ACM-Reference-Format}
\bibliography{shortenedbiblio}

\end{document}